\documentclass[useAMS,usenatbib]{mn2e}

\usepackage{times,graphicx,amssymb,natbib}

\title[Convergence of SPH disc simulations]{Convergence of SPH simulations of self-gravitating accretion discs: \\
Sensitivity to the implementation of radiative cooling}
\author[W.K.M. Rice, D.H. Forgan and P.J. Armitage]{W.K.M. Rice$^{1}$\thanks{E-mail:
wkmr@roe.ac.uk}, D.H. Forgan$^{1}$, and P.J. Armitage$^{2,3}$\\
$^{1}$Scottish Universities Physics Alliance (SUPA), Institute for Astronomy, University of Edinburgh, Blackford Hill, Edinburgh, EH9 3HJ, UK \\
$^{2}$JILA, 440 UCB, University of Colorado, Boulder, CO 80309-0440, USA \\
$^{3}$Department of Astrophysical and Planetary Sciences, University of Colorado, Boulder, USA}
\begin{document}

\date{Accepted 0000}

\pagerange{\pageref{firstpage}--\pageref{lastpage}} \pubyear{0000}

\maketitle

\label{firstpage}

\begin{abstract}
Recent simulations of self-gravitating accretion discs, carried out using a three-dimensional 
Smoothed Particle Hydrodynamics (SPH) code by Meru and Bate, have been interpreted as 
implying that three-dimensional global discs fragment much more easily than would be expected 
from a two-dimensional 
local model. Subsequently, global and local two-dimensional models have been shown to display similar fragmentation 
properties, leaving it unclear whether the three-dimensional results reflect a physical effect 
or a numerical problem associated with the treatment of cooling or artificial viscosity in SPH. 
Here, we study how fragmentation of self-gravitating disc flows in SPH depends upon the 
implementation of cooling. We run disc simulations that compare a simple cooling scheme, in 
which each particle loses energy based upon its internal energy per unit mass, with a method 
in which the cooling is derived from a smoothed internal energy density field. For the 
simple per particle cooling scheme, we find a significant increase in the minimum cooling 
time scale for fragmentation with increasing resolution, matching previous results. Switching 
to smoothed cooling, however, results in lower critical cooling time scales, and tentative 
evidence for convergence at the highest spatial resolution tested. We conclude that precision 
studies of fragmentation using SPH require careful consideration of how cooling (and, probably, 
artificial viscosity) is implemented, and that the apparent non-convergence of the fragmentation 
boundary seen in prior simulations is likely a numerical effect. In real discs, where cooling 
is physically smoothed by radiative transfer effects, the fragmentation boundary is probably 
displaced from the two-dimensional value by a factor that is only of the order of unity.
\end{abstract}


\section{Introduction}

\noindent
Accretion discs exist in many astrophysical systems - in x-ray binaries and cataclysmic 
variable stars, around supermassive black holes in the nuclei of active galaxies, 
and around young, newly forming stars.  In some cases, in particular in active
galactic nuclei and around young stars, these discs can be sufficiently
massive that their own self-gravity plays an important role in their evolution. 
The susceptibilty of an infinitesimally thin accretion disc to the growth of the gravitational instability 
can be determined using the Toomre $Q$ parameter \citep{toomre64}
\begin{equation}
Q = \frac{c_s \Omega}{\pi G \Sigma},
\label{eq:Q}
\end{equation}
where $c_s$ is the disc sound speed, $\Omega$ is the angular frequency, $G$ is the 
gravitational constant, and $\Sigma$ is the disc surface mass density.  Discs are
unstable to axisymmetric perturbations if $Q < 1$, while numerical simulations
\citep{durisen07} suggest that the growth of non-axisymmetric perturbations occurs
for $Q < 1.5 - 1.7$. 

It has been understood for quite some time that the gravitational instability
can act to transport angular momentumm outwards, allowing mass to accrete onto the
central object \citep{lin87, laughlin94, armitage01} and could well be the primary
transport mechanism during the earliest stages of star formation \citep{lin87,rice10}. 
It is also possible that, if sufficiently unstable, these discs may fragment to form bound objects \citep{kuiper51}. 
In the case of protostellar discs, these objects could contract to form gas giant
planets \citep{boss98,boss00} or could form stars in discs around supermassive black holes 
\citep{shlosman89,goodman03, bonnell08}.

\citet{paczynski78} suggested that non-fragmenting, self-gravitating discs may exist in
a state of marginal stability. \citet{gammie01}, using two-dimensional, shearing sheet simulations, quantified this by showing that, in addition
to the \citet{toomre64} parameter, the evolution of a self-gravitating accretion disc is
also controlled by the rate at which it loses energy.  Using a cooling time, $\tau_c$, of the following form
\begin{equation}
\tau_c = \beta \Omega^{-1},
\label{eq:betacool}
\end{equation}
where $\beta$ is a constant, \citet{gammie01} showed that if $\beta > 3$, the system settles into a quasi-steady, marginally
stable state.  If not, the system fragments to form bound objects.  Consistent results were obtained by \citet{rice03} using
three-dimensional smoothed particle hydrodynamic (SPH) simulations, although the actual cooling 
time boundary was found to depend on the chosen equation of state \citep{rice05}.

Since self-gravitating discs transport angular momentum outwards, it is useful to describe them as
having an effective viscosity.  \citet{shakura73} suggested that an appropriate form 
for the viscosity in an accretion disc is
\begin{equation}
\nu = \alpha c_s H,
\label{eq:shakura}
\end{equation}
where $\alpha << 1$ is the viscosity parameter and $H$ is the disc thickness.  \citet{gammie01}
showed that the cooling time in a self-gravitating accretion disc could be related
to the viscosity through an effective gravitational $\alpha$ that satisfies
\begin{equation}
\alpha_{\rm eff} = \frac{4}{9 \gamma (\gamma - 1) \beta},
\label{eq:alpha_eff}
\end{equation}   
where $\gamma$ is the specific heat ratio.  It has since been shown \citep{cossins09} that the amplitude of
the perturbations in such a disc is related to $\alpha_{\rm eff}$ (or $\beta$) through
\begin{equation}
\frac{<\delta \Sigma>}{<\Sigma>} \propto \alpha_{\rm eff}^{1/2} \propto \frac{1}{\sqrt{\beta}}.
\label{eq:perturb}
\end{equation} 
Essentially the current understanding is that the amplitude of the perturbations increase
with increasing $\alpha_{\rm eff}$ and, hence, there is a maximum value for $\alpha_{\rm eff}$
that a self-gravitating disc can sustain without fragmenting. This maximum has been shown to be $\alpha_{\rm max} \sim 0.06$
\citep{gammie01, rice05}.

\citet{meru11} have, however, recently shown that simulations of self-gravitating discs using SPH
fail to converge.  In their simulations, the $\beta$ value at which
fragmentation occurs increases as the resolution increases. They do infer that this may be a numerical effect, but also
suggest that this could mean that fragmentation may occur for arbitrarily long cooling times.  
If fragmentation requires short cooling times, then it becomes very unlikely that it can occur in the inner regions
of protostellar discs \citep{rafikov05,boley06,stamatellos08,clarke09} and, hence, is no longer a viable mechanism
for the formation of gas giant planets.  On the other hand, if fragmentation can occur for very long cooling times
it would mean that gas giants planets could form via gravitational collapse. However, this would be somewhat
surprising as it would imply that fragmentation could occur for infinitesimally small
perturbation amplitudes.  We suggest, in this paper, that it is indeed a numerical effect and is related
to the manner in which cooling is typically implemented in SPH.  

That the results in \citet{meru11} are likely to  be numerical
suggests that our understanding of self-gravitating accretion discs is effectively unchanged. 
In a typical isolated disc that has a relatively low mass compared to the mass of the central star, 
the evolution is largely determined by $Q$ and the cooling time.  If $Q \sim 1$ and the cooling time is such that
$\alpha_{\rm eff} < 0.06$, it settles into a quasi-steady, marginally stable state with relatively
small perturbations and in which
angular momentum is transported outwards.  If $\alpha_{\rm eff} > 0.06$ the perturbations become
non-linear and fragmentation occurs.  

This paper is organised as follows. In Section 2 we describe SPH and how the implementation of the
cooling may introduce numerical effects.  In Section 3 we describe our results and show that implementing
cooling in a manner that is more consistent with the SPH formalism may lead to convergence.  In Section 4 we
discuss these results and reach a few conclusions.

\section{Smoothed Particle Hydrodynamics}
\subsection{The Basic Formalism}
Smoothed Particle Hydrodynamics (SPH) is a Lagrangian hydrodynamic formalism in which
a fluid is represented by pseudo-particles (see, for example, \citealt{benz90,monaghan92}). 
Each particle in the simulation has a mass ($m$), position ($x,y,z$), velocity ($v_x,
v_y, v_z$) and internal energy per unit mass ($u$).

Although SPH uses particles, the interpretation is that each particle represents a
smeared out distribution of density.  The contribution that particle $j$, located
at $\textbf{r}_j$, makes to the density at location $\textbf{r}$ is
\begin{equation}
\rho_j(\textbf{r}) = m_j W(|\textbf{r} - \textbf{r}_j|;h_j),
\label{eq:jdens}
\end{equation}
where $W$ is the ``smoothing kernel'' that describes the form of the mass distribution and
$h_j$ is the smoothing length associated with particle $j$.  The density at $\textbf{r}$ 
is then determined by summing over the $N$ particles that contribute to the 
density at that location
\begin{equation}
\rho(\textbf{r}) = \sum_{j=1}^N \rho_j(\textbf{r})
\label{dens}
\end{equation} 

To perform a basic hydrodynamic simulation, SPH needs an equation of state and has to solve the continuity equation,
the momentum equation and the energy equation.  If the energy equation is included then typically an ideal gas equation
of state will be used
\begin{equation}
P_j = (\gamma - 1) \rho u_j
\label{eq:eos}
\end{equation}
where $\gamma$ is the specific heat ratio, $\rho$ is the fluid density at the location of
particle $j$, and $u_j$ is the internal energy per unit mass associated with particle $j$.

The continuity equation is
\begin{equation}
\frac{\partial \rho}{\partial t} + \nabla \cdot{(\rho \textbf{v})} = 0,
\label{eq:continuity}
\end{equation}
but since SPH uses particles with constant
masses, mass is conserved trivially and the continuity equation does not need to
be solved explicitly.  Ignoring self-gravity, the Lagrangian form 
of the momentum equation is
\begin{equation}
\frac{d \textbf{v}}{d t} = -\frac{1}{\rho}\nabla{P},
\label{eq:momentum}
\end{equation} 
showing that - in the absence of gravity - the acceleration is determined only by the local pressure gradient.  
In SPH, the acceleration of each particle is determined and Equation (\ref{eq:momentum}) 
is typically rewritten into a suitable form shown below \citep{benz90,monaghan92}
\begin{equation}
\frac{d \textbf{v}_j}{d t} = -\sum_i m_i \left(\frac{P_j}{\rho_j^2} + \frac{P_i}{\rho_i^2} \right) \nabla_i W(|\textbf{r}_i - \textbf{r}_j|;h),
\label{eq:SPHmomentum}
\end{equation}
where the sum is over all the neighbours ($i$) of particle $j$ and $\nabla_i W(|\textbf{r}_i - \textbf{r}_j|;h)$ is the gradient of the kernel 
(a known function) at the location of each neighbour, with $h$ typically the average of the smoothing lengths for particles $i$ and $j$.  

The energy equation is
\begin{equation}
\frac{\partial e}{\partial t} + \nabla \cdot(e \textbf{v}) = -P \nabla \cdot \textbf{v},
\label{eq:energy}
\end{equation} 
where $e = \rho u$ is the internal energy density.  Using the continuity equation (Equation (\ref{eq:continuity}))
the energy equation can be written in its Lagrangian form
\begin{equation}
\frac{d u}{d t} = -\frac{P}{\rho} \nabla \cdot \textbf{v}
\label{eq:LagrangianEnergy}
\end{equation}
which can then also be written into a form more suitable for SPH (e.g., \citet{monaghan92})
\begin{equation}
\frac{d u_j}{d t} = \frac{1}{2} \sum_i \left( \frac{P_i}{\rho_i^2} + \frac{P_j}{\rho_j^2} \right) \textbf{v}_{ij} \cdot \nabla_i W(|\textbf{r}_i - \textbf{r}_j|;h),
\label{eq:SPHEnergy}
\end{equation}
where $\textbf{v}_{ij}$ is the velocity of particle $i$ with respect to particle $j$.

\subsection{Introducing cooling}
To investigate how self-gravitating discs evolve in the presence of cooling, many simulations \citep{gammie01,rice03} 
have used what is now referred to as $\beta$-cooling and is illustrated in Equation (\ref{eq:betacool}).  The cooling term in the energy
equation would therefore be
\begin{equation}
\left( \frac{d e}{d t} \right)_{\rm cool} = -\frac{e}{\tau_c}.
\label{eq:CoolingTerm}
\end{equation} 

If this is introduced into the energy equation, Equation (\ref{eq:LagrangianEnergy}) then becomes
\begin{equation}
\frac{d u}{d t} = -\frac{P}{\rho} \nabla \cdot \textbf{v} - \frac{e}{\rho}\frac{1}{\tau_c} = -\frac{P}{\rho} \nabla \cdot \textbf{v} - \frac{u}{\tau_c}. 
\label{eq:LagEnerCool}
\end{equation}
Since each particle has an internal energy per unit mass ($u_j$) associated with it, typically the cooling term for 
each particle is simply assumed to be
\begin{equation}
\left( \frac{d u_j}{d t} \right)_{\rm cool} = -\frac{u_j}{\tau_c}.
\label{eq:CoolTermParticle}
\end{equation} 
This cooling form has been used in many previous SPH simulations (e.g., \citealt{rice03,rice05,cossins09}) and is the form
used by \citet{meru11} in their simulations that appear not to converge as the resolution is increased.

The internal energy per unit mass associated with a particle is, however, the total amount of thermal energy that the particle has and, in an SPH sense - as with the mass - 
should be regarded as a smeared out distribution of thermal energy.  We argue that using the standard $\beta$-cooling formalism  produces a mismatch in scales.  The hydrodynamic equations are solved at the scale of the smoothing length $h$ - if the gravitational potential is kernel-softened, then it can be argued that gravitational forces are also limited by the smoothing length $h$.  For simulations that are not well-resolved, a density peak of length-scale $\lambda$ will be suppressed by a factor of $1 + h/\lambda$, as noted by \citet{lodato11}, who use this to derive a resolution requirement for fragmentation:

\begin{equation} 
\beta_{\rm res} = \beta_0\left(1 + h/H\right)^{-2}.
\end{equation}

In implementing $\beta$-cooling at the locations of particles only, the cooling operates far below the $h$-scale.  Purely hydrodynamic simulations \emph{will} calculate $u$ in a smoothed sense, but adding an unsmoothed $\beta$-cooling term violates this.  We should then seek a means by which the $\beta$-cooling term can be limited to the $h$-scale.  In SPH, any quantity, $A$, associated with the fluid can be
estimated using density-weighted interpolations between the known particle values (e.g., \citet{bodenheimer07})
\begin{equation}
A(\textbf{r}) = \sum_{i=1}^N m_i \frac{A_i}{\rho_i} W(|\textbf{r}-\textbf{r}_j|, h).
\label{eq:Interpolate}
\end{equation}
Formally, the internal energy
density at the location of particle $j$ - at position $\textbf{r}$ - should then be determined using
\begin{equation}
e(\textbf{r}) = \rho(\textbf{r}) \sum_i \frac{m_i}{\rho_i} u_i W(|\textbf{r}-\textbf{r}_i|,h).
\label{eq:SmoothEnergy}
\end{equation}
The internal energy per unit mass at the location of particle $j$ is therefore, formally, $e(\textbf{r})/\rho(\textbf{r})$.  
In principle, one would expect
that - at the location of each particle - the interpolated value should be the same as the value assigned to the
particle.  This is the case in any regions that vary smoothly, but is not the case at shocks and other discontinuities
\citep{brownlee07}.  That fragmentation tends to occur in the spiral shock regions may, therefore, suggest that using the unsmoothed
thermal energies per unit mass may result in errors in the cooling rate that could be removed by using the smoothed
thermal energy.  We should acknowledge that
there is a potential inconsistency in that the gas pressure is typically defined using Equation (\ref{eq:eos}) in which
a smoothed quantity ($\rho$) is combined with an unsmoothed quantity ($u$). As suggested by \citet{ritchie01}, it may be more consistent to define the
pressure as
\begin{equation}
P_j = (\gamma - 1) \rho(r) \sum_{i} \frac{m_i}{\rho_i} u_i W(|\textbf{r}_j-\textbf{r}_i|,h).
\label{eq:eos_mod}
\end{equation}
However, the pressure force and heating terms in Equations (\ref{eq:SPHmomentum}) and (\ref{eq:SPHEnergy}) are determined via
interpolation and so it is not clear that the pressure itself has to be determined via interpolation.

If the above is correct, then it is possible that the non-convergence seen in \citet{meru11} is due to
this mismatch between $u_j$ and the smoothed internal energy per unit mass at the location of
particle $j$.  We could simply replace Equation (\ref{eq:CoolTermParticle}) with
\begin{equation}
\left( \frac{d u_j}{d t} \right)_{\rm cool} = -\frac{1}{\tau_c}\sum_i \frac{m_i}{\rho_i} u_i W(|\textbf{r}_j-\textbf{r}_i|,h).
\label{eq:SmoothCool1}
\end{equation}
This, however, has a problem because the cooling is exponential and, hence, any particle cooler than all its neighbours 
will very quickly (and artifically) cool to a very low thermal energy.  Instead, we distribute the cooling, associated with 
particle $j$, 
across the neigbour sphere and use
\begin{eqnarray}
\left( \frac{d u_j}{d t} \right)_{\rm cool} &=& -\frac{1}{\tau_c} \frac{W(0,h_j) m_j u_j}{\rho_j} \nonumber \\
\left( \frac{d u_{i \ne j}}{d t} \right)_{\rm cool} &=& - \frac{1}{\tau_c} \frac{m_i}{\rho_i} u_i W(|\textbf{r}_j-\textbf{r}_i|,h),
\label{eq:SmoothCool2}
\end{eqnarray}
where the lower of the two equations in Equation (\ref{eq:SmoothCool2}) is applied to all neighbours $i$ of particle $j$.
Ultimately, the total cooling rate for particle $j$ will be
\begin{equation}
\left( \frac{d u_j}{d t} \right)_{\rm cool} = -\frac{1}{\tau_c}\sum_i \frac{m_j}{\rho_j} u_j W(|\textbf{r}_j-\textbf{r}_i|,h),
\label{eq:totalcoolj}
\end{equation}
which ensures that the correct amount of energy is removed from the neighbour sphere, but prevents the coolest particle from
cooling exponentially to an artificially low thermal energy. We suggest here that using Equation (\ref{eq:totalcoolj}) to represent $\beta$-cooling is more appropriate than
using Equation (\ref{eq:CoolTermParticle}) as it more correctly represents the cooling rate at the location of
particle $j$ and removes the correct amount of energy per time from the neighbour sphere, which is the smallest scale we can expect the SPH formalism to correctly accommodate. 

\subsection{Shock-tube test}\label{sec:shock-tube}
As discussed above, we suggest that the non-convergence seen in the
\citet{meru11} simulations could be a consequence of their implementation of $\beta$-cooling. As with many others
(e.g., \citealt{rice03}, \citealt{rice05}, \citealt{cossins09}) the cooling rate is determined using the internal energy
per unit mass assigned to each particle.  We propose that it is more appropriate to use a smoothed thermal energy
(e.g., Equation (\ref{eq:SmoothEnergy})) to determine the cooling rate. 

However, the interpolation used in SPH (e.g., Equation (\ref{eq:Interpolate})) should, when applied at the location of a
particle, recover the value of the fluid quantity assigned to that particle.  If so, one would expect that smoothed
cooling and basic cooling should produce the same results.  There are, however, regions - in particular at discontinuities such as shock waves or contact 
discontinuities - where this is not the case (see for example \citealt{brownlee07}). As a simple test we considered a one-dimensional shock tube test 
problem (e.g., \citealt{hernquist89}) which we carried out using SPH and
using the grid-based {\small{PENCIL CODE}} \citep{brandenburg03}. 

\begin{figure}
\begin{center}
\includegraphics[scale = 0.45]{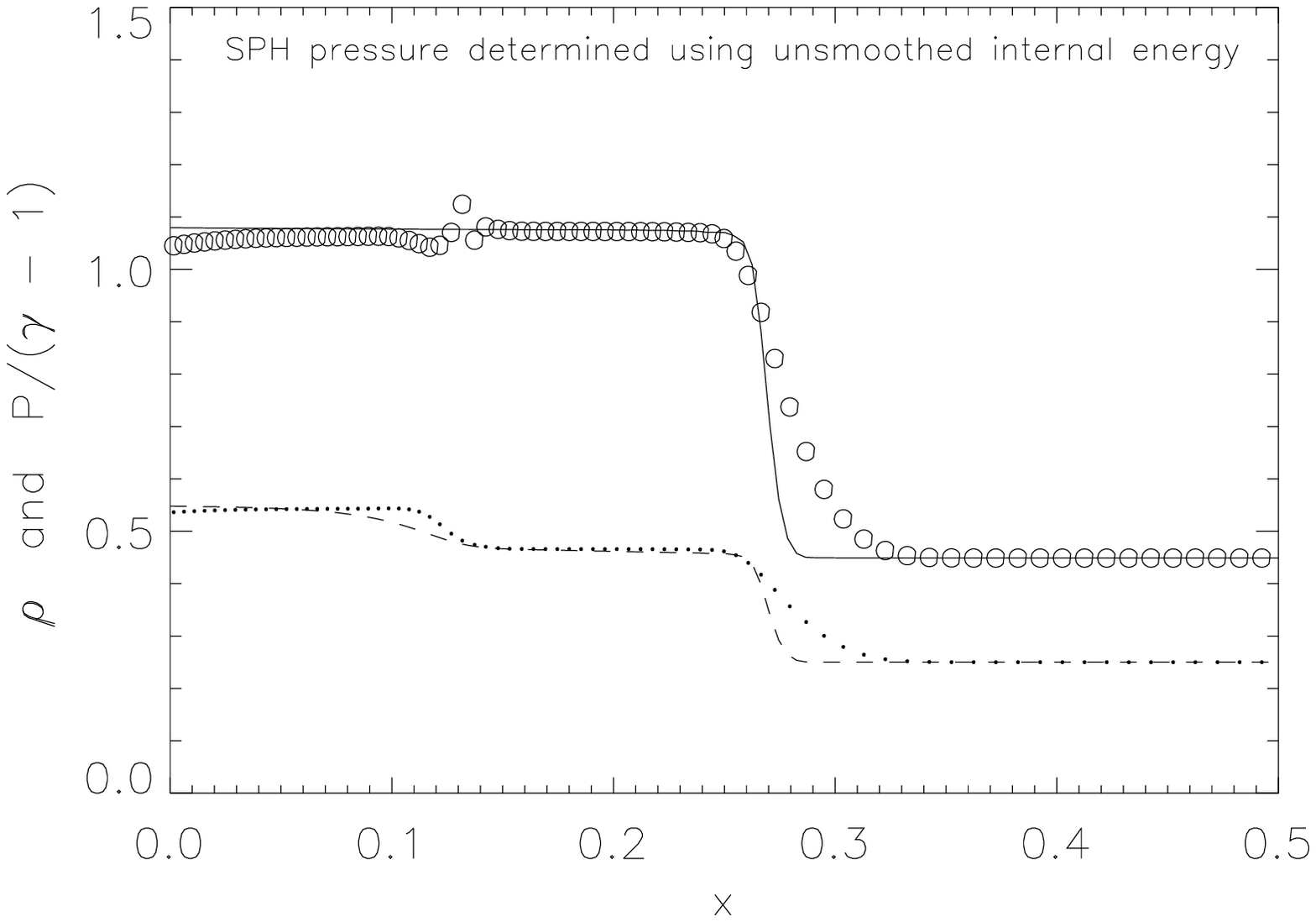}
\includegraphics[scale = 0.45]{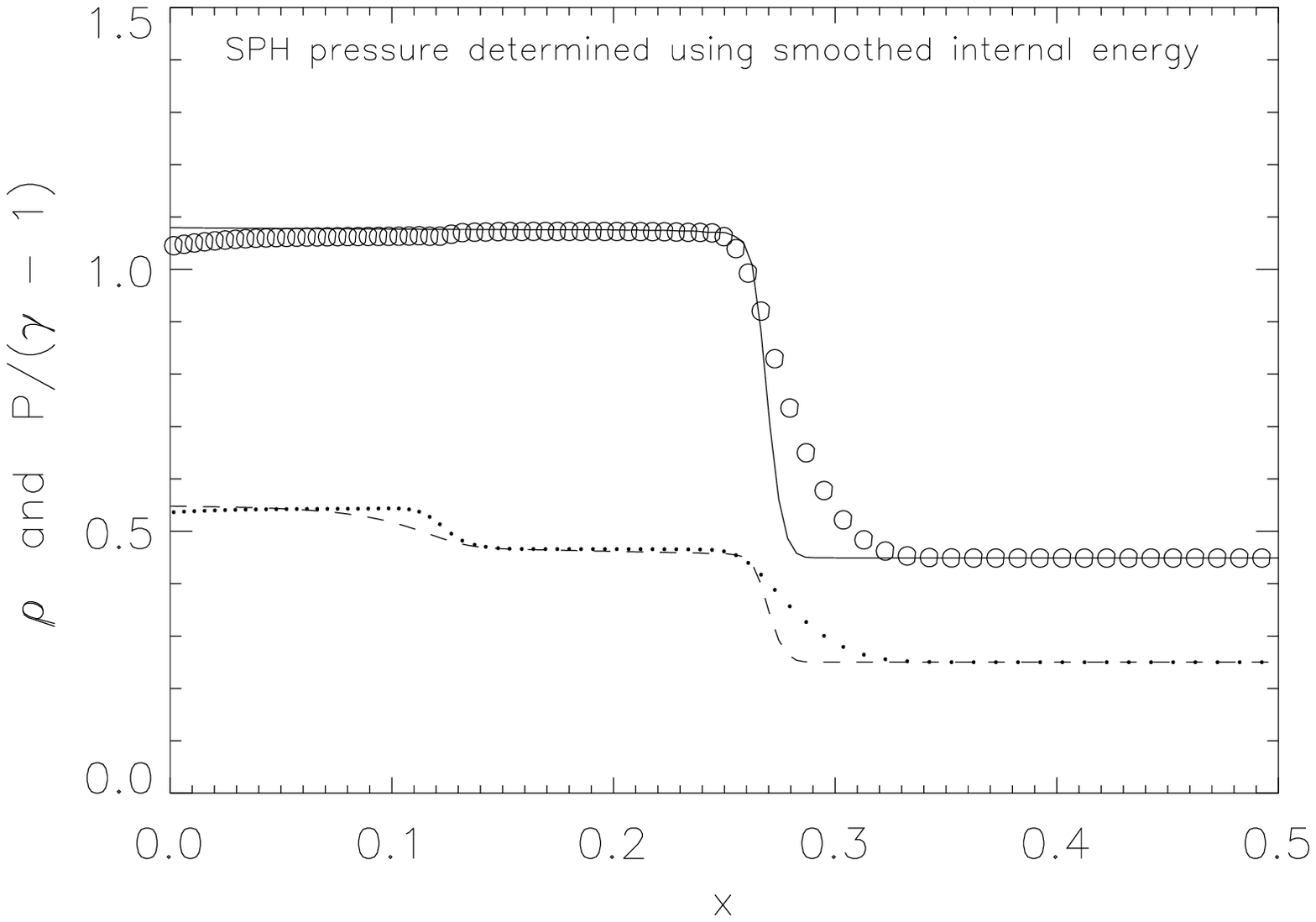}
\caption{Comparison between an SPH shock tube test problem (symbols) and one performed using the \small{PENCIL CODE} (lines). 
These show a shock wave at $x \sim 0.3$ and a contact discontinuity at $x \sim 0.1$. In the top panel,
the SPH pressure (open circles) is determined using the unsmoothed internal energy resulting in an unphysical jump
at the contact discontinuity.  Using the smoothed internal energy (lower panel) removes this unphysical
discontinuity in the pressure.}
\label{fig:comparison}
\end{center}
\end{figure}

Figure \ref{fig:comparison} shows the density, $\rho$, (dashed line) and pressure $(P = (\gamma - 1) \rho u)$ divided by $(\gamma - 1)$ (solid line) from the 
{\small{PENCIL CODE}} compared
with the corresponding values from the SPH simulation (symbols).  The pressure is divided by $(\gamma - 1)$ simply to seperate it from the 
density.  In the top panel it is clear that at the contact discontinuity (located at $x \sim 0.1$), the {\small{PENCIL CODE}} pressure 
is continuous (as expected) while the SPH pressure - determined using the unsmoothed 
internal energy per unit mass - is not. This discontinuity in the SPH pressure at the contact discontinuity is unphysical but,
since the pressure force (Equation (\ref{eq:SPHmomentum})) and energy term (Equation (\ref{eq:SPHEnergy})) are determined by
interpolation, the SPH code does not behave as if this discontinuity is actually present. If the cooling,
however, uses the unsmoothed internal energy, then this discontinuity will be present in the cooling and will 
produce an unphysical mismatch between cooling and heating near the contact discontinuity. Since fragmentation
typically occurs in the region behind the shock, this mismatch could indeed play an important role in promoting 
fragmentation.  The lower panel in Figure \ref{fig:comparison} shows that the discontinuity in pressure at the
contact discontinuity is not present if the smoothed internal energy is used to determine the pressure.  This suggests
that using the smoothed internal energy in the cooling formalism is more physically correct than using the 
unsmoothed internal energy.  We therefore, below,
carry out a series of simulations to determine if convergence is achieved when Equation (\ref{eq:totalcoolj})
is used instead of Equation (\ref{eq:CoolTermParticle}).

\section{Results}
All of the simulations presented have the same basic conditions as \citet{meru11}, which are based on those of \citet{lodato04}.  A central
star with a mass of $M_{\rm star} = 1$ surrounded by a disc extending from $r_{\rm in} = 0.25$ to
$r_{\rm out} = 25$, with a mass of $M_{\rm disc} = 0.1$ and with a surface density profile of
$\Sigma \propto r^{-1}$.  These are, as with \citet{meru11}, scale-free simulations and so
we don't explicitly assume any scale, although typically these could be regarded as representing
a disc from $0.25$ to $25$ au around a star with a mass of $1$ M$_\odot$.   

\subsection{Comparison with earlier simulations}
To ensure that our SPH code is comparable to that of \citet{meru11} we first run simulations using 
the basic cooling as described by Equation
(\ref{eq:CoolTermParticle}). Table \ref{tab:sims1} shows the simulations that we performed using basic cooling. 
You'll notice that, unlike \citet{meru11}, we do not carry out any 31250 particle simulations.  As shown by
\citet{bate97}, an SPH simulation only correctly represents fragmentation if the minimum resolvable mass
\begin{equation}
M_{\rm min} = 2 M_{\rm tot} \frac{N_{\rm neigh}}{N_{\rm tot}}
\label{eq:MinMass}
\end{equation}
is less than the Jeans mass.  In Equation (\ref{eq:MinMass}) $M_{\rm tot}$ is the total disc mass, 
$N_{\rm neigh}$ is the number of neighbours each particle has 
(typically 50) and $N_{\rm tot}$ is the total number of particles in the simulation.

\begin{figure}
\begin{center}
\includegraphics[scale = 0.45]{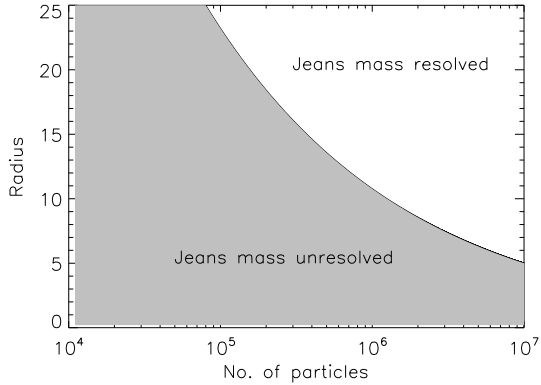}
\caption{Figure showing, for an SPH simulation with $\Sigma \propto r^{-1}$, $M_{\rm disk} = 0.1$, $M_{\rm star} = 1$, and 
$r_{\rm out} = 25$, the radial range where the Jeans mass is resolved (unfilled) and unresolved (filled) plotted against particle number. This clearly shows that
if fewer than $80000$ particles are used, the Jeans mass is not properly resolved anywhere in the disc.}
\label{fig:JeansResol}
\end{center}
\end{figure}

The Jeans mass can be approximated using $M_J = \pi \Sigma H^2$ where $H = c_s/\Omega$ is the disc scale height. If we
use $\Sigma = \Sigma_o/r$ then, given that $M_{\rm disc} = 0.1$, $\Sigma_o = 0.1/(2 \pi [r_{\rm out} - r_{\rm in}]) \approx 0.1/(2 \pi r_{\rm out})$. Since these
discs are self-gravitating, we can use $Q = c_s \Omega / (\pi G \Sigma) = 1$ \citep{toomre64} to give
\begin{equation}
c_s = \frac{\pi G \Sigma}{\Omega} = \pi \Sigma_o r^{1/2},
\label{eq:cs}
\end{equation}
where we've also used that $M_{\rm star} = 1$ and $G = 1$. We can therefore show that the Jeans mass in these 
simulations has the following dependence on radius
\begin{equation}
M_J = \frac{0.001 r^3}{8 r_{\rm out}^3}.
\label{eq:JeansMass}
\end{equation}
Figure \ref{fig:JeansResol} shows a plot of radius against particle number where the filled region shows the radial range for which the
Jeans mass is not resolved while the unfilled region shows the radial range for which the Jeans mass is resolved. Figure \ref{fig:JeansResol}
shows clearly that if fewer than $80000$ particles are used, the Jeans mass is not resolved anywhere in the disc.
With 250000 particles, the Jeans mass is resolved beyond $r = 17$ and for 2 million particles it is resolved beyond $r = 8$. That the
\citet{meru11} results include simulations with 31250 particle, and that these simulations are consistent with the resulting trend,
suggests that numerical effects are influencing the results.  We also don't carry out any simulations
with more than 2 million particles simply because our code is parallelised using OpenMP and we are therefore limited as to how many
processors we can access and, hence, can't realistically carry out higher resolution simulations.

\begin{table}
\centering
\begin{minipage}{140mm}
  \caption{List of the simulations using basic cooling. \label{tab:sims1}}
  \begin{tabular}{c || ccc}
  \hline
  \hline
   Simulation & No. of particles &  $\beta$ & Fragment?  \\
 \hline
  1 & 250000 & 4 & Yes  \\
  2 & 250000 & 5 & Yes  \\
  3 & 250000 & 6 & No  \\
  4 & 250000 & 7 & No  \\
  5 & 500000 & 5 & Yes  \\
  6 & 500000 & 7 & Yes \\
  7 & 500000 & 8 & No  \\
  8 & 2000000 & 8 & Yes \\ 
 \hline
  \hline
\end{tabular}
\end{minipage}
\end{table}

\begin{figure}
\begin{center}
\includegraphics[scale = 0.5]{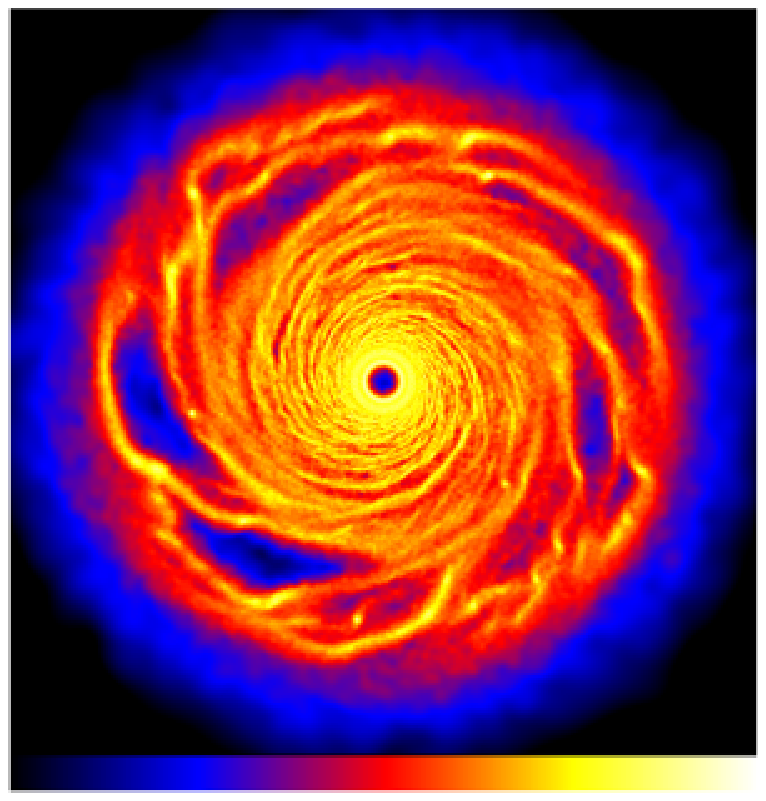} 
\includegraphics[scale=0.5]{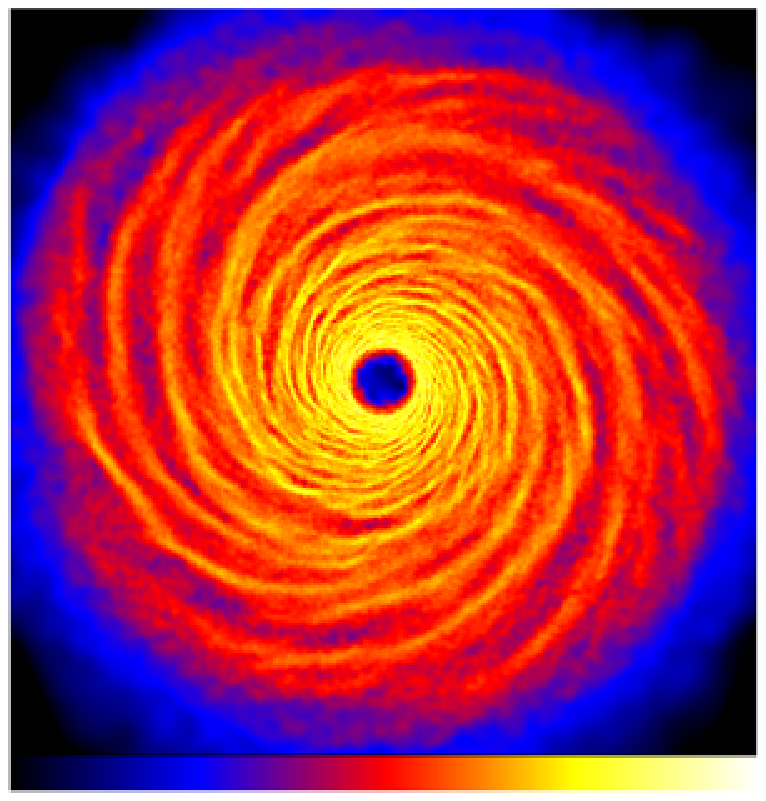} 
\caption{Simulations with 250000 particles and using basic cooling.  The left panel is
$\beta=4$ while the right panel is $\beta = 6$.}
\label{fig:250Basic}
\end{center}
\end{figure}

\begin{figure}
\begin{center}
\includegraphics[scale = 0.5]{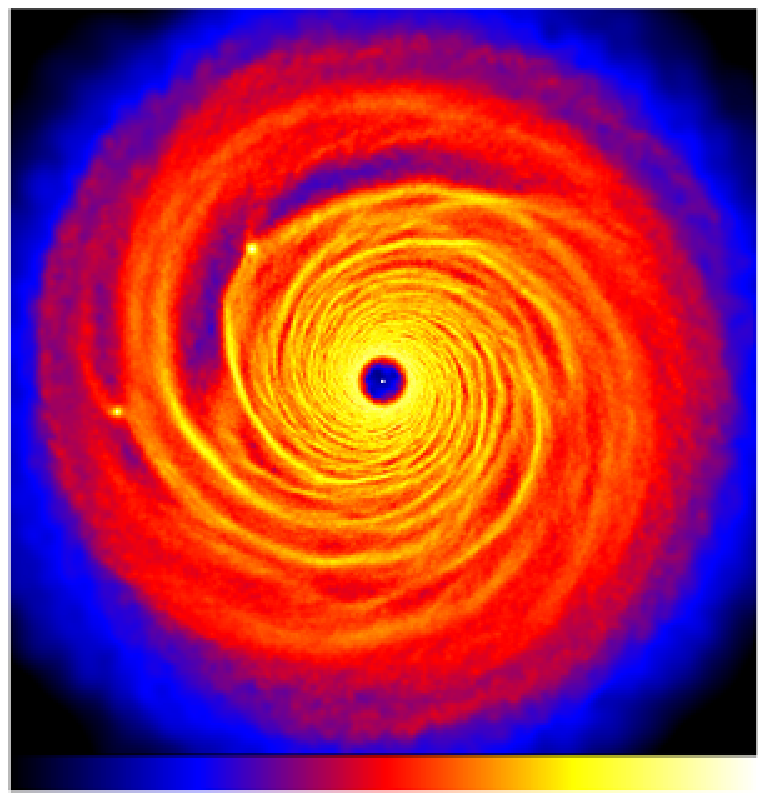}
\includegraphics[scale = 0.5]{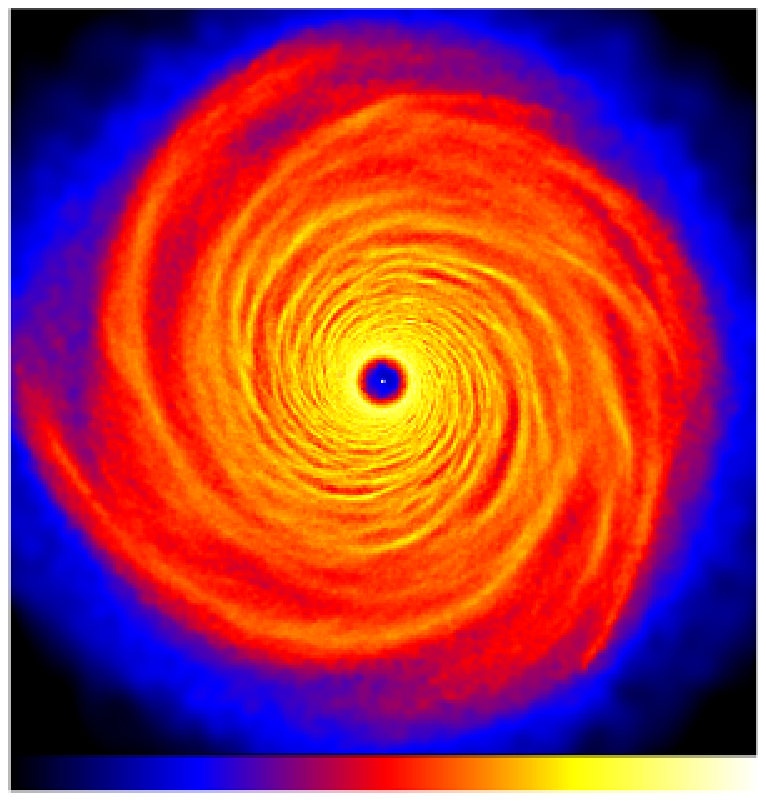}
\caption{Simulations with 500000 particles and using basic cooling.  Left panel is
$\beta=7$ and the right panel is $\beta = 8$.}
\label{fig:500Basic}
\end{center}
\end{figure}

\begin{figure}
\begin{center}
\includegraphics[scale = 0.5]{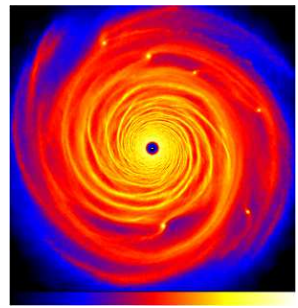}
\caption{Simulation with 2 million particles, using basic cooling and with
$\beta = 8$.}
\label{fig:2mBasic}
\end{center}
\end{figure}

Table \ref{tab:sims1} shows that these simulations are consistent with those of \citet{meru11}.  For 250000 particles,
fragmentation occurs for $\beta$ between 5 and 6, for 500000 particles it is between 7 and 8 and for 2 million
particles fragmentation occurs for $\beta > 8$.  Figs. \ref{fig:250Basic}, \ref{fig:500Basic}, and \ref{fig:2mBasic}
show surface density images of a number of the simulations.  For the 250000 and 500000 particle simulations 
we show an image of simulation that fragmented and one that did not.  For the 2 million particle simulation we simply
illustrate (in Fig. \ref{fig:2mBasic}) that the $\beta = 8$ simulation did indeed undergo fragmentation. These simulations
are all run for at least 6 outer rotation periods and, if there is any evidence for fragmentation, they are continued until
either these fragments become much denser than the local density, or they shear away. The clumps
in Figs. \ref{fig:250Basic}, \ref{fig:500Basic} and \ref{fig:2mBasic} are therefore smaller and fewer in number than
would be the case if the image showed an epoch just after fragmentation started.

\subsection{Smoothed cooling}
Having shown that, using the basic cooling represented by Equation (\ref{eq:CoolTermParticle}), we
get results consistent with \citet{meru11} we then carried out a large number of simulations
using what we shall refer to as smoothed cooling (Equation (\ref{eq:totalcoolj})).  Table \ref{tab:sims2}
lists all of these simulations and whether or not fragmentation occured.  The starred 10 million
particle simulations are ones in which 4 million particles (with a total mass of $M_{\rm disc} = 0.04$) 
were placed in a ring from $r = 15$ to $r = 25$.  This is equivalent to 10 million particles (with
a total mass of $M_{\rm disc} = 0.1$) placed between $r = 1$ and $r = 25$, and so we refer to these
as 10 million particle simulations. 

\begin{table}
\centering
\begin{minipage}{140mm}
  \caption{List of the simulations using smoothed cooling. \label{tab:sims2}}
  \begin{tabular}{c || ccc}
  \hline
  \hline
   Simulation & No. of particles &  $\beta$ & Fragment?  \\
 \hline
  1 & 250000 & 4 & Yes  \\
  2 & 250000 & 4.5 & Yes \\
  3 & 250000 & 5 & No  \\
  4 & 250000 & 6 & No  \\
  5 & 250000 & 7 & No  \\
  6 & 500000 & 5 & Yes  \\
  7 & 500000 & 6 & Yes \\
  8 & 500000 & 7 & No  \\
  9 & 500000 & 8 & No \\
  10 & 2000000 & 5 & Yes \\
  11 & 2000000 & 6 & Yes \\
  12 & 2000000 & 7 & No \\
  13 & 2000000 & 8 & No \\
  14 & 10000000$^*$ & 7 & Yes \\
  15 & 10000000$^*$ & 8 & Yes \\
  16 & 10000000$^*$ & 9 & No \\
 \hline
  \hline
\end{tabular}
\end{minipage}
\end{table}

\begin{figure}
\begin{center}
\includegraphics[scale = 0.5]{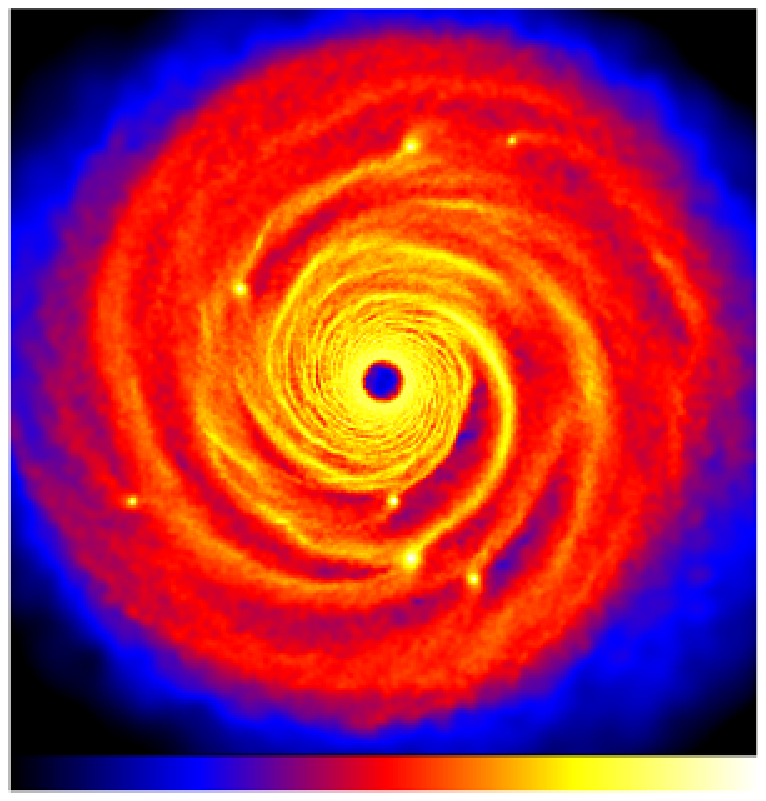}
\includegraphics[scale = 0.5]{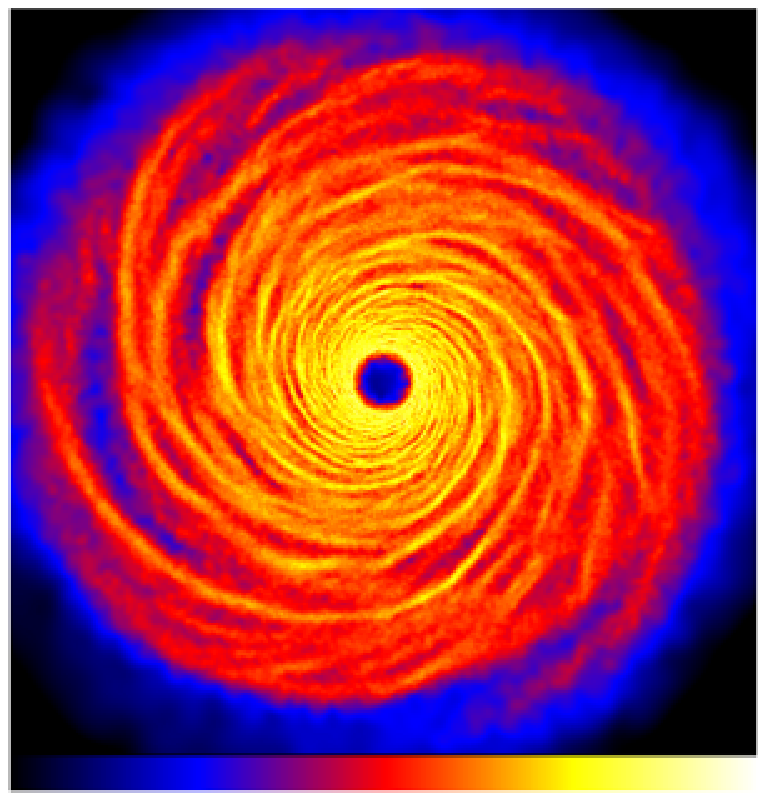}
\caption{Simulations with 250000 particles and using smoothed cooling.  The left panel is
$\beta=4$, and the right panel is $\beta = 5$.}
\label{fig:250Smooth}
\end{center}
\end{figure}

\begin{figure}
\begin{center}
\includegraphics[scale = 0.5]{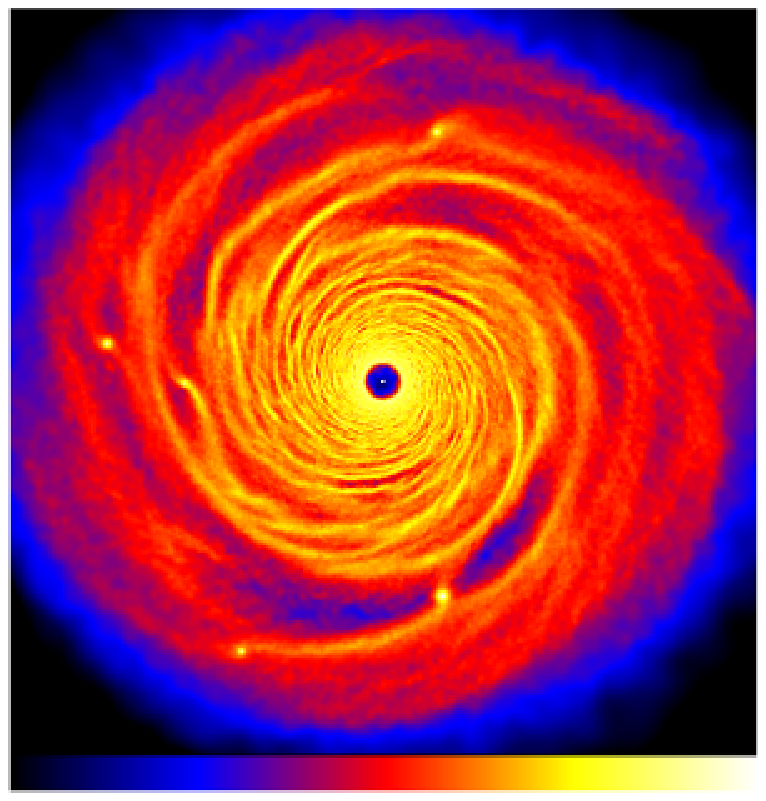}
\includegraphics[scale = 0.5]{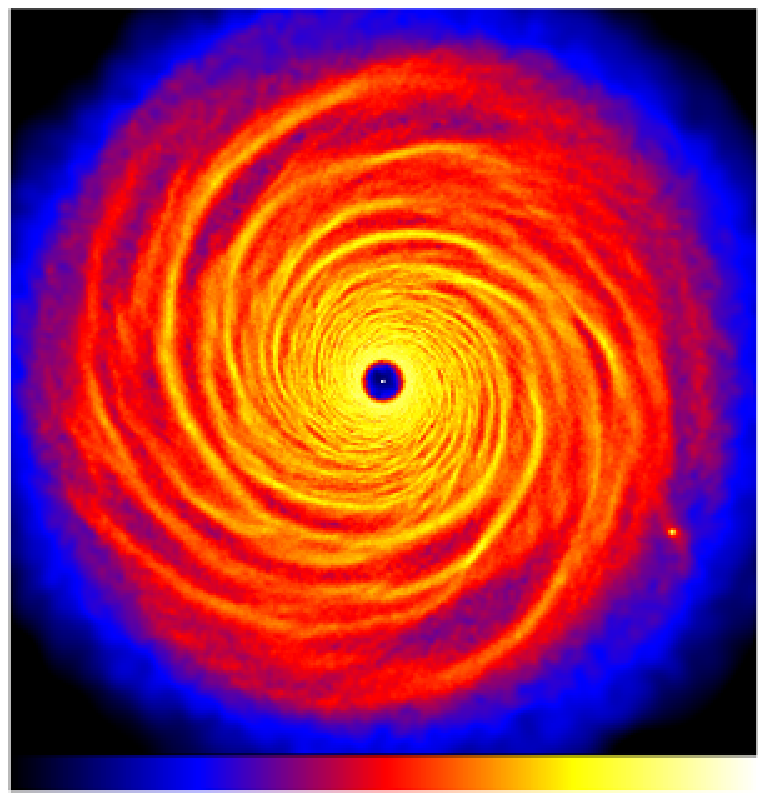}
\includegraphics[scale = 0.5]{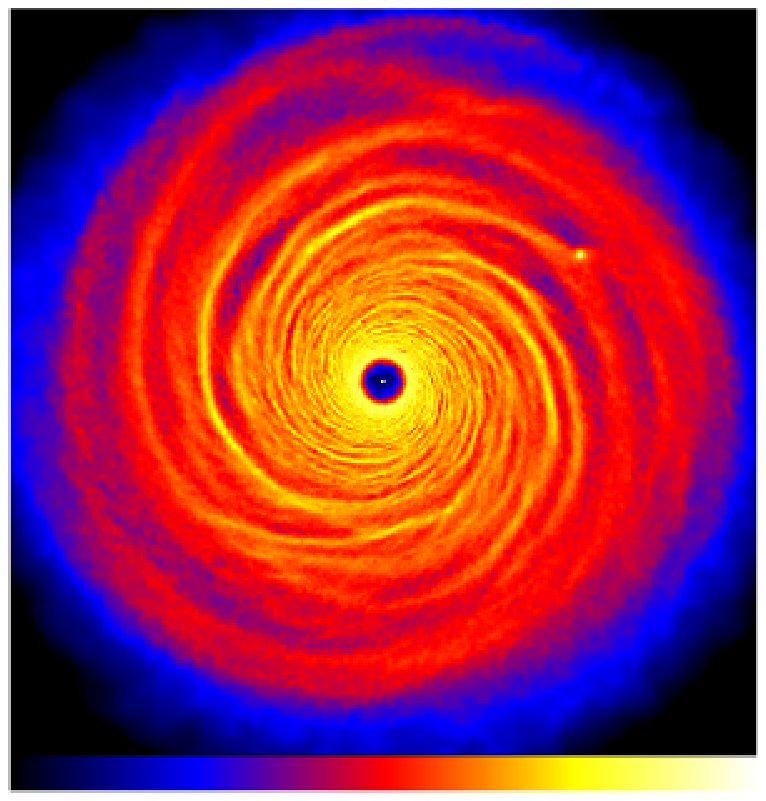}
\includegraphics[scale = 0.5]{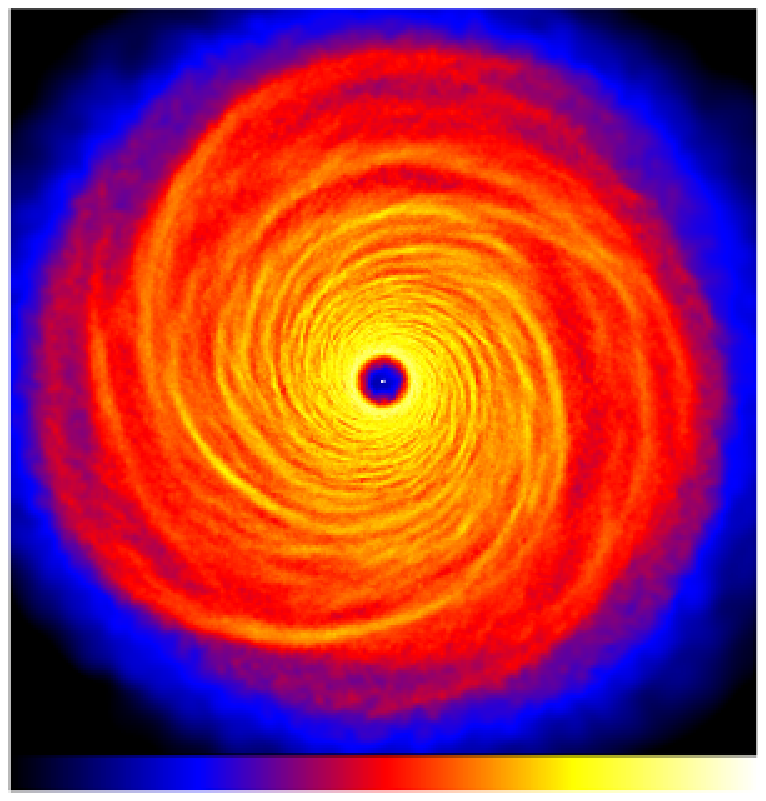}
\caption{Simulations with 500000 particles and using smoothed cooling.  The $\beta$ values are
$\beta = 4$ (top left), 5 (top right), 6 (bottom left), and 7 (bottom right).}
\label{fig:500Smooth}
\end{center}
\end{figure}

\begin{figure}
\begin{center}
\includegraphics[scale = 0.5]{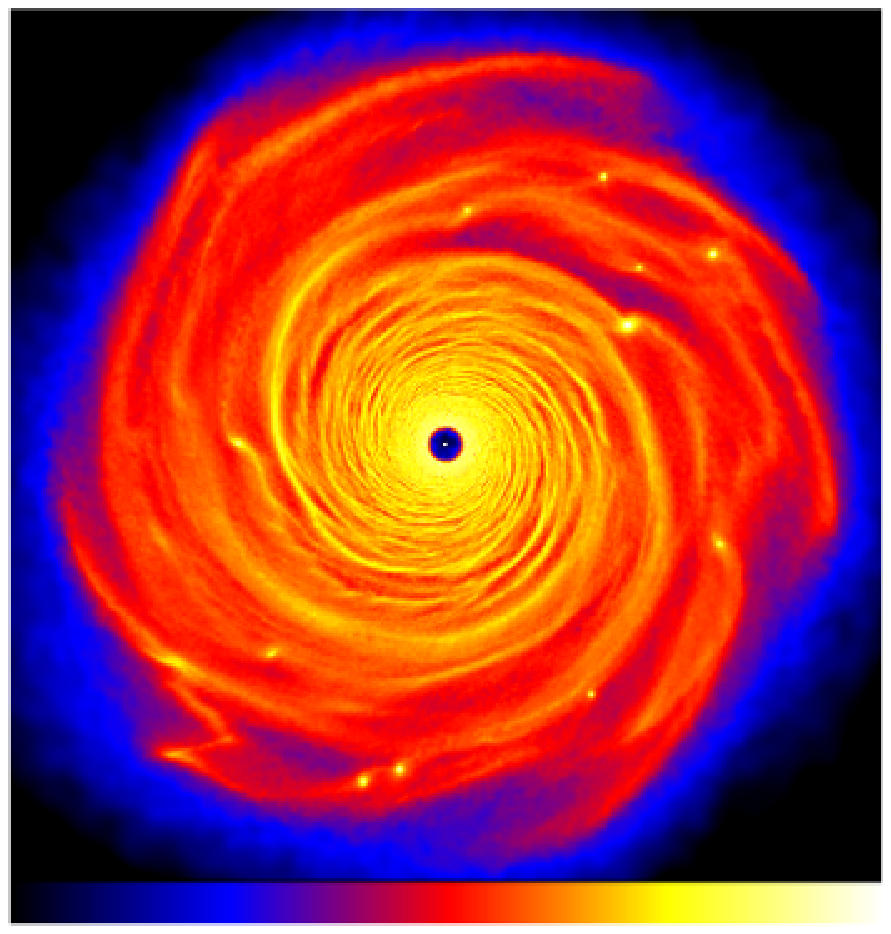}
\includegraphics[scale = 0.5]{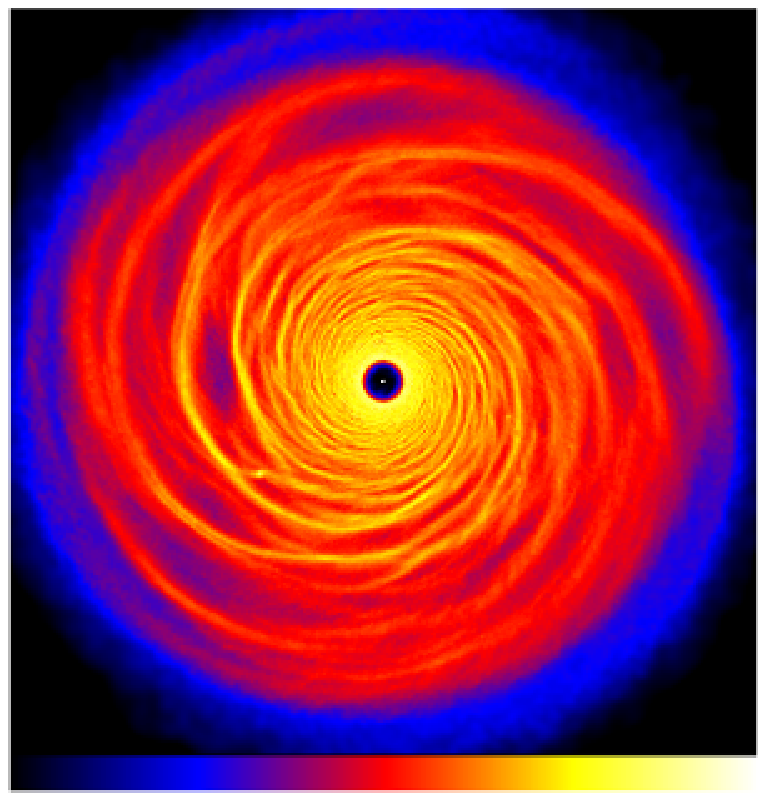}
\includegraphics[scale = 0.5]{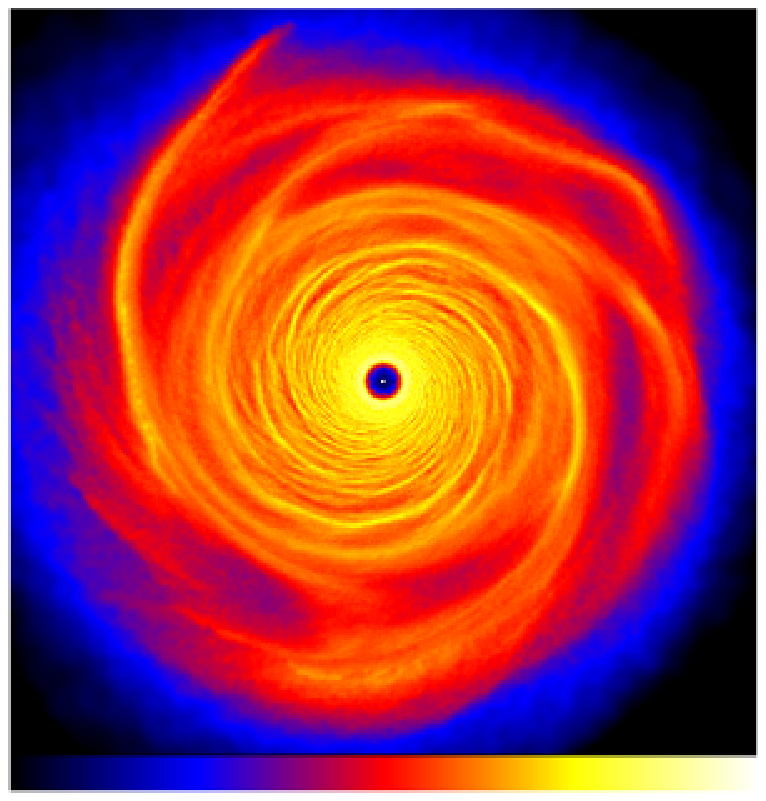}
\caption{Simulations with 2 million particles and using smoothed cooling.  The top-left panel is
$\beta=5$, the top right-panel is $\beta=6$ and the bottom panel is $\beta = 7$.}
\label{fig:2mSmooth}
\end{center}
\end{figure}

\begin{figure}
\begin{center}
\includegraphics[scale = 0.1]{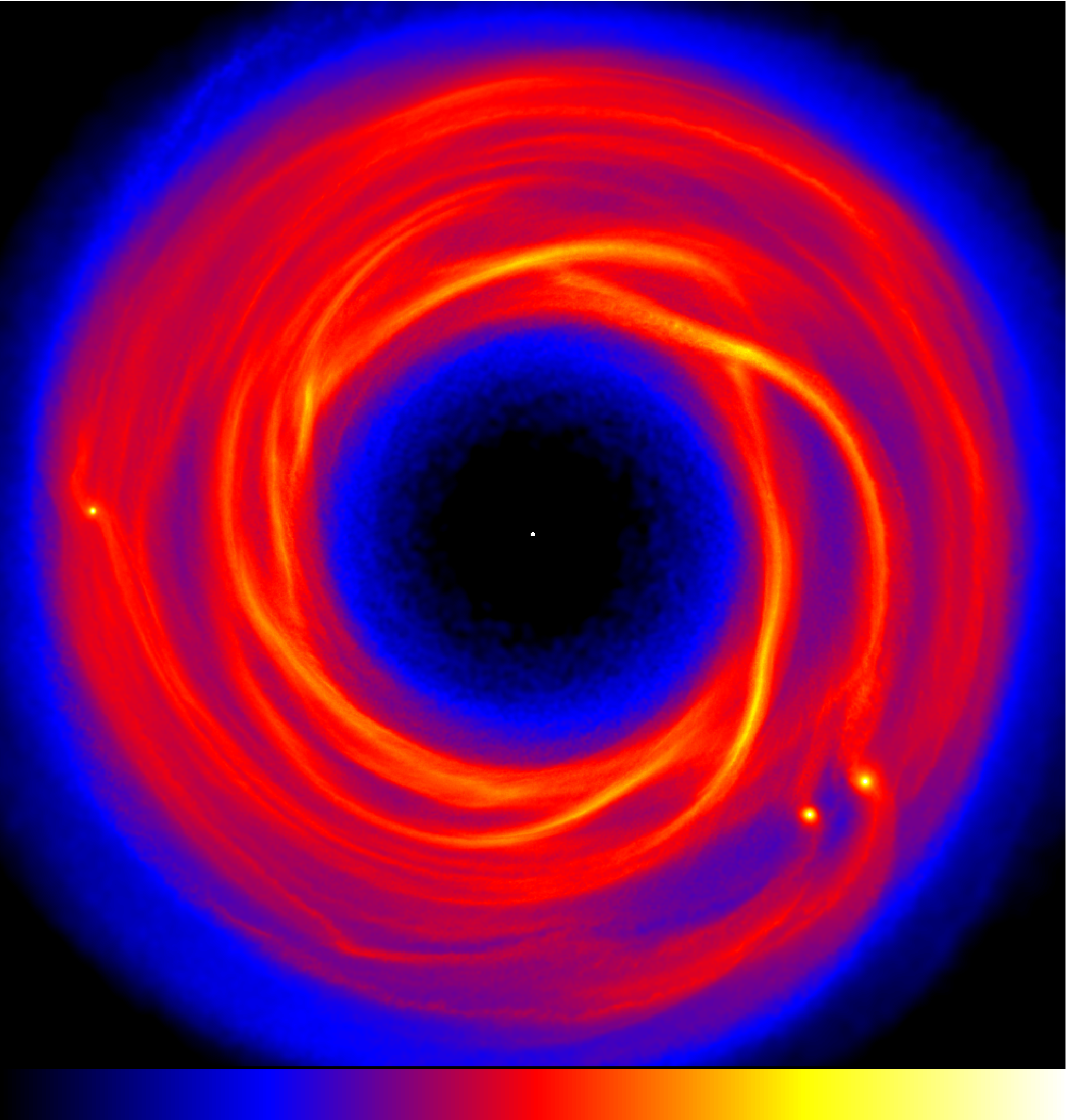}
\includegraphics[scale = 0.1]{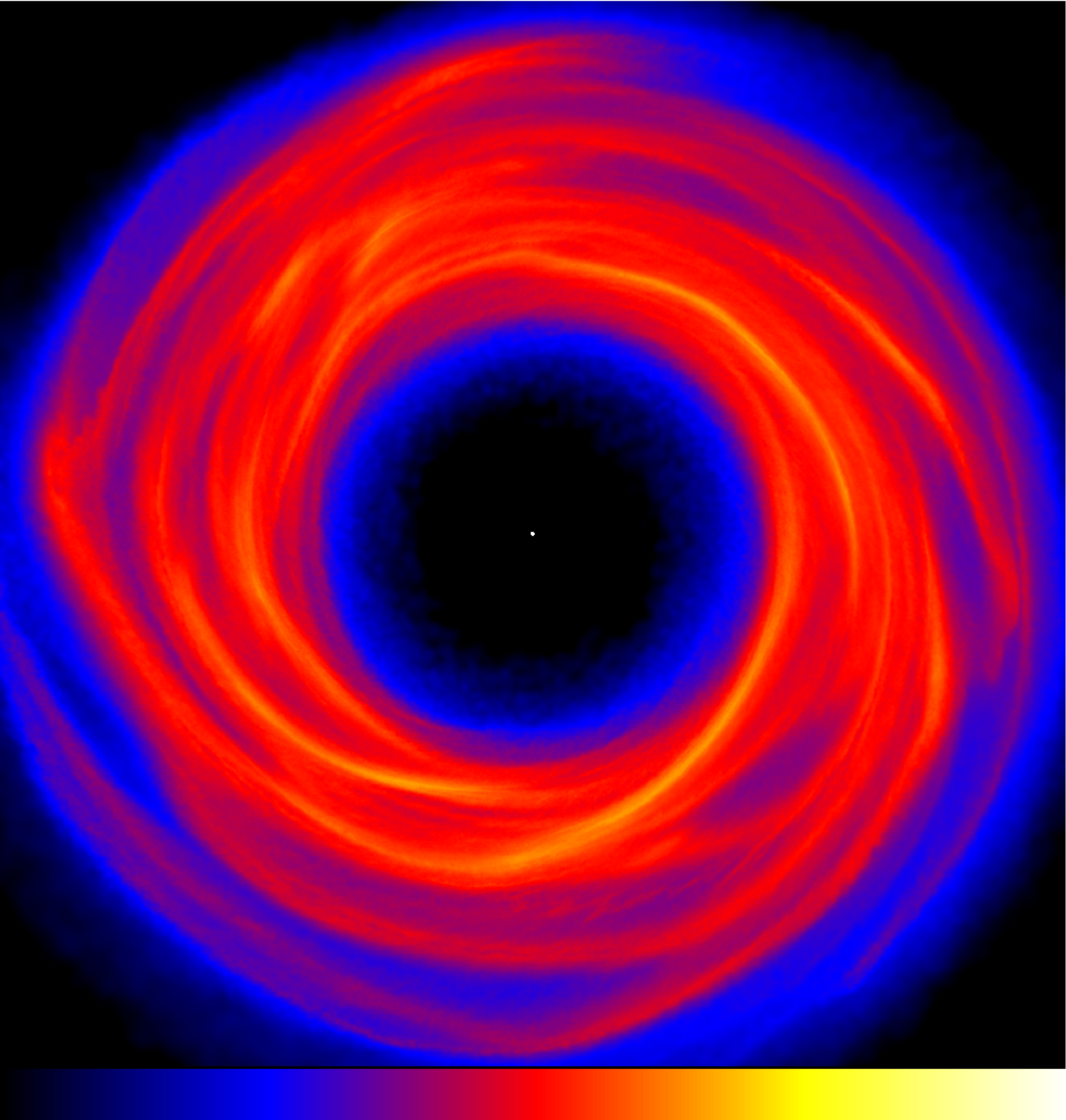}
\caption{Simulations with a ring consisting of 4 million particles and using smoothed cooling. This is
equivalent to a disc simulation with 10 million particles. Left panel is
$\beta=8$ and the right panel is $\beta=9$.}
\label{fig:10mSmooth}
\end{center}
\end{figure}

Figs. \ref{fig:250Smooth}, \ref{fig:500Smooth}, \ref{fig:2mSmooth} and \ref{fig:10mSmooth} show surface
density images from the simulations using smoothed cooling showing the range of 
$\beta$ values across which the simulations go from fragmenting to non-fragmenting.

\begin{figure}
\begin{center}
\includegraphics[scale = 0.5]{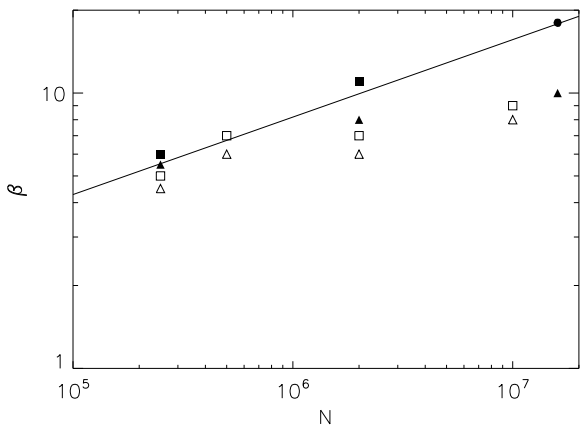}
\caption{Figure showing fragmentation boundary from Meru \& Bate (filled symbols) together with results from this work (open symbols).
The line represent the relationship between $\beta_{\rm crit}$ and particle number determined by \citet{meru11}.}
\label{fig:FragBound}
\end{center}
\end{figure}

The results from \citet{meru11} suggest that the dependence of the fragmentation boundary ($\beta_{\rm crit}$) on
resolution is $\beta_{\rm crit} \propto n^{1/3}$ with $n$ the number of particles in the simulation. This is 
equivalent to saying that the fragmention boundary depends linearly on resolution (i.e., $h \propto n^{1/3}$).
Table \ref{tab:sims2} immediately shows that our results are considerably different to those of \citet{meru11}
with the fragmentation boundary varying from between 4.5 and 5 for simulations with 250000 particles to
between 8 and 9 for the simulation with effectively 10 million particles.  Fig. \ref{fig:FragBound}
compares the results from \citet{meru11} with the results obtained here. The filled symbols are from
\citet{meru11} and the open symbols are from this work.  Triangles represents simulations that fragmented
while squares are for those that did not.  The filled circle represents a \citet{meru11} simulation that
they referred to as being borderline.  We see very little evidence for what might be regarded as borderline
systems. Our simulations either definitely fragment (the clumps continue to grow and become extremely dense) 
or the system settles into a quasi-steady state with no signs of fragmentation. In some cases clumps will form and 
shear away, but this appears to happen very rarely.  We would generally interpret the \citet{meru11} borderline cases
as non-fragmenting systems. The line in Fig. \ref{fig:FragBound} shows the relationship between 
$\beta_{\rm crit}$ and particle number determined by \citet{meru11}.

It is clear from Fig. \ref{fig:FragBound} that using smoothed cooling (Equation (\ref{eq:totalcoolj}))
produces results significantly different to that produced by simulations using basic cooling (Equation (\ref{eq:CoolTermParticle})).  
Although the effective 10 million particle simulation has a fragmentation boundary slightly higher than
the 2 million and 500000 particle simulations, it does appear as though we can conclude that the
fragmentation boundary may be converging to a value of $\beta_{\rm crit} < 10$.  That the 500000 particle
and 2 million particles simulations produced very similar values for $\beta_{\rm crit}$ (between $\beta  = 6$
and $\beta = 7$) had lead us to initially conclude that these simulations had already converged.  The higher
value for $\beta_{\rm crit}$ produced by the 10 million particle simulation does suggest that convergence
hasn't yet been reached, but could also imply that a ring of particles is not an exact proxy for a more
massive and larger disc with more particles. 

\subsection{Comparison of cooling rates}
\begin{figure}
\begin{center}
\includegraphics[scale = 0.5]{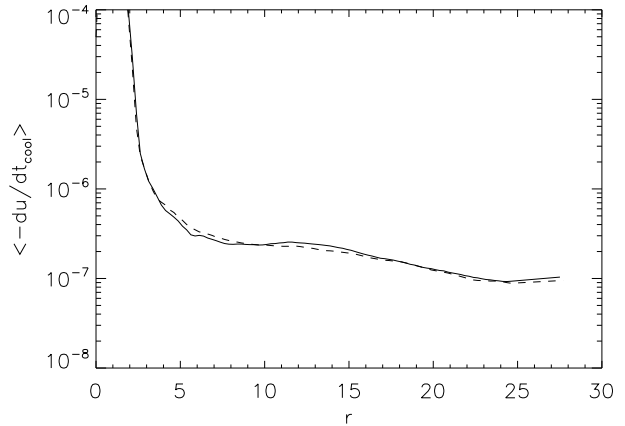}
\caption{Comparison of the cooling rate using smoothed cooling (solid line) and basic cooling (dashed line).  These cooling rates
are azimuthally averaged and also averaged over the final 1.5 outer rotation periods. The inner regions of the disc is dominated
by numerical viscosity which artificially heats this inner region.  Beyond $r = 5$, there can be local variations that are as large
as 10 \%, but globally the two cooling formalisms are very similar, differing by only a few percent.}
\label{fig:dudt_cool}
\end{center}
\end{figure}

That the interpolated value of the thermal energy should approximate well the individual particle values everywhere except at
discontinuities suggests that there shouldn't be a significant difference, in general, between the cooling rate in simulations
using basic cooling when compared to those using smoothed cooling.  To test this we considered azimuthally
averaged cooling rates for 2 of the simulations with the same number of particles (500000) and both with $\beta = 8$ that
did not fragment when either smoothed or basic cooling was used.  We also averaged over the final 10 dumps covering a time
of 1.5 outer rotation periods.  The result is shown in Figure \ref{fig:dudt_cool} with the solid line being the cooling rate
using smoothed cooling and the dashed line being the cooling rate using basic cooling.   

Figure \ref{fig:dudt_cool} show that the cooling rates are indeed similar and have the same radial trend.  The large increase in 
cooling rate inside $r = 5$ is a consequence
of this region being poorly resolved due to particles having been accreted onto the central star \citep{bate95}, resulting
in the numerical viscosity dominating in this inner region (see for example \citealt{forgan11}). This artificially heats the inner disc and produces the large increase
in cooling rate.  Beyond $r = 5$, where numerical viscosity does not dominate and the system is well resolved, the local variations 
can, however, be as large as $10$ \%.  If, however, we sum the individual particle cooling rates from a single dump and take the average 
for $r > 5$, the smoothed and basic cooling formalisms differ by, typically, less than 5 \% with neither cooling formalism being, consistently,
the most efficient.  If this is averaged across the final 1.5 outer rotation periods (10 dumps) the difference is only 1.8 \%. This suggests
that even though locally there can be reasonable (10 \%) differences between the two cooling formalisms, globally they are very similar.

\section{Discussion and conclusions}    
To study the evolution of self-gravitating accretion discs that are undergoing cooling,
many authors have used a form of cooling known as $\beta$-cooling and which is represented by
Equation (\ref{eq:betacool}). In SPH, this has generally been implemented by cooling
each particle with a cooling rate determined using the internal energy per unit
mass associated with that particle. \citet{meru11} have recently shown, however,
that such simulations do not converge and that the value for $\beta$ at
which fragmentation occurs ($\beta_{\rm crit}$) increases linearly with resolution
(i.e., $\beta_{\rm crit} \propto n^{1/3}$).

We confirm here that this non-convergence is indeed seen in such simulations, but suggest that it 
may be a SPH problem and may result from
using the internal energy per unit mass associated with each particle.  This presents a mismatch of scales between the 
$h$-scale (where the traditional SPH equations operate) and the essentially infinitesimal scales at which 
the additional $\beta$-cooling term operates.  We advocate an alternative approach to 
$\beta$-cooling where the cooling rate is determined by smoothing over contributions due to the neighbours. This
correctly represents the local internal energy per unit mass, and removes any unphysical discontinuities 
(in pressure for example) that could occur in the shocked regions when using the unsmoothed internal energy. 
Although our simulations using this smoothed cooling do not strictly converge, they appear to be converging 
to $\beta_{\rm crit} < 10$ (for $\gamma = 5/3$), consistent with results from earlier work. To further test the
convergence of our proposed cooling method would require higher resolution simulations ($ > 10$ million particles)
that we are not able, at this stage, to carry out in a reasonable time.  We also compare
the cooling rates using these two cooling formalisms.  Although there are local variations that can be as
high as $10$ \%, globally (if we consider the regions beyond $r = 5$, that are well resolved and not dominated by numerical
viscosity) the two cooling methods differ by only a few percent.  This is consistent with our suggestion that
it is local differences in the two cooling methods (occuring primarily at shocks and other discontinuities) that  
result in different fragmentation boundaries.

It has also been suggested \citep{paardekooper11} that this non-convergence is
due to the use of smooth initial conditions.  Fragmentation then occurs at the boundary
between the turbulent and laminar regions as the system cools towards a quasi-steady
state.  It is suggested that it may be more appropriate to produce an initial quasi-steady
state using a long cooling time (a high $\beta$ value) and then to vary the value of $\beta$.  
This is a very reasonable suggestion but comparisons of their work with this paper may not be wholly appropriate.  Their implementation of $\beta$-cooling in a 2D grid is closer to our smoothed cooling techniques than traditional SPH $\beta$-cooling, which (given the results of this paper) would assist efforts to converge.  On the other hand, their simulations require a seed of white noise to be added to the smooth initial conditions to generate spiral structure.  SPH simulations typically contain Poisson noise in the initial conditions as a result of innate particle disorder, and the resulting fragmentation is relatively insensitive to the level of this noise \citep{cartwright09}.  These differing noise properties may provide a reason as to why grid-based simulations converge more easily after being relaxed.  In any case, we argue that their results are not sufficient to explain the linear dependence on resolution seen by \citet{meru11}.  

Although we have focused on simulations that use $\beta$-cooling, the results presented here also have potential implications for 
SPH simulations that approximate radiation transfer (e.g., \citealt{whitehouse04, mayer07, stamatellos07, forgan09}).
In most of these formalisms, the temperature at the location of a particle is determined from its internal energy per unit mass. It is possible that it too should
be determined by smoothing across the neighbour sphere.  The form of the cooling in these radiation transform formalisms is, however, very different to that for $\beta$-cooling and typically these
simulations cool towards an equilibrium temperature rather than exponentially towards zero (as is the case for $\beta$-cooling), so the convergence problem seen with
$\beta$-cooling may not affect these simulations.

Ultimately what we suggest here is that the non-convergence seen by \citet{meru11} is primarily
a numerical effect and hence that there is no evidence to suggest that fragmentation can occur
for arbitrarily long cooling times in self-gravitating accretion discs.  The standard picture that
fragmentation requires short cooling times ($\alpha_{\rm eff} > 0.06$) still, therefore, appears
to be valid.  This is consistent with the results of \citet{cossins09} and \citet{rice11} who show that the perturbation
amplitudes in self-gravitating discs depend inversely on the cooling time and hence fragmentation occurs
when the perturbation are large (non-linear) and, consequently, when the cooling time is short. 

\section*{Acknowledgements}

\noindent 
All simulations presented in this work were carried out using high performance computing funded by the Scottish Universities Physics
Alliance (SUPA). KR and DF gratefully acknowledge support from STFC grant ST/H002380/1. PJA ackowledges support 
from NASA, under awards NNX09AB90G and NNX11AE12G from the Origins of Solar Systems and 
Astrophysics Theory programs, and from the NSF under award AST-0807471.  The authors would like to thank
Cathie Clarke, David Hubber and Dan Price for useful discussions.

\label{lastpage}


\begin{thebibliography}{}

\bibitem[Armitage, Livio \& Pringle(2001)]{armitage01} 
 Armitage P.J., Livio M., Pringle J.E., 2001, MNRAS, 324, 705

\bibitem[Artymowicz \& Lubow(1994)]{artymowicz94}
 Artymowicz P., Lubow S.H., 1994, ApJ, 421, 651

\bibitem[Bate, Bonnell \& Price(1995)]{bate95}
 Bate M.R., Bonnell I.A., Price N.M., 1995, MNRAS, 277, 362

\bibitem[Bate \& Burkert(1997)]{bate97}
 Bate M.R., Burkert A., 1997, MNRAS, 288, 1060

\bibitem[Benz(1990)]{benz90}
 Benz W., 1990, in Buchler J.R., ed., Numerical Modelling of Nonlinear Stellar Pulsations Problems and Prospects. Kluwer, Dordrecht, p. 269

\bibitem[Bodenheimer et al.(2007)]{bodenheimer07}
 Bodenheimer P., Laughlin G.P., R\'ozycka M., Yorke H.W., eds, 2007, Numerical Methods in Astrophysics: An Introduction. Taylor Francis, New York

\bibitem[Boley et al.(2006)]{boley06}
 Boley A.C., Mejia A.C., Durisen R., Cai K., Pickett M.K., D'Alessio P., 2006, ApJ, 651, 517

\bibitem[Bonnell \& Rice(2008)]{bonnell08}
 Bonnell I.A., Rice W.K.M., 2008, Science, 321, 1060

\bibitem[Boss(1998)]{boss98}
 Boss A.P., 1998, Nat., 393, 141

\bibitem[Boss(2000)]{boss00}
 Boss A.P., 2000, ApJ, 536, L101

\bibitem[Brandenburg(2003)]{brandenburg03}
 Brandenburg A., 2003, in Advances in Nonlinear Dynamos, ed. A. Ferriz-Mas
\& M. N\'u\~nez (London: Taylor \& Francis), 269

\bibitem[Brownlee et al.(2007)]{brownlee07}
 Brownlee R.A., Houston P., Levesley J., Rosswog S., 2007, in Algorithms for Approximation: 
Proceedings of the 5th International Conference, Chester UK, p. 103

\bibitem[Cartwright et al.(2009)]{cartwright09}
Cartwright A., Stamatellos D., Whitworth A., 2009, MNRAS, 395, 2373

\bibitem[Clarke(2009)]{clarke09}
 Clarke C.J., 2009, MNRAS, 396, 1066

\bibitem[Cossins, Lodato \& Clarke(2009)]{cossins09}
 Cossins P., Lodato G., Clarke C.J., 2009, MNRAS, 393, 1157

\bibitem[Durisen et al.(2007)]{durisen07}
 Durisen R., Boss A.P., Mayer L., Nelson A.F., Quinn T., Rice W.K.M., 2007,
 in Reipurth B., Jewitt D., Keil K., eds, Protostars and Planets V, Gravitational
 Instabilities in Gaseous Protoplanetary Disks and Implications for 
 Giant Planet Formation, University of Arizona Press

\bibitem[Forgan et al.(2009)]{forgan09}
 Forgan D., Rice K., Stamatellos D., Whitworth A., 2009, MNRAS, 394, 882

\bibitem[Forgan et al.(2011)]{forgan11}
 Forgan D., Rice K., Cossins P., Lodato G., 2011, MNRAS, 410, 994

\bibitem[Gammie(2001)]{gammie01}
 Gammie C.F., 2001, ApJ, 553, 174

\bibitem[Goodman(2003)]{goodman03}
 Goodman J., 2003, MNRAS, 339, 937

\bibitem[Herquist \& Katz(1989)]{hernquist89}
 Hernquist L., Katz N., 1989, ApJS, 70, 419

\bibitem[Kuiper(1951)]{kuiper51}
 Kuiper G., 1951, in Hynek J.A., ed., Proceedings of a topical symposium, c
commemorating the 50th anniversary of the Yerkes Observatory and half a century
of progress in astrophysics, McGraw-Hill, New York, p. 357

\bibitem[Laughlin \& Bodenheimer(1994)]{laughlin94}
 Laughlin G., Bodenheimer P., 1994, ApJ, 436, 335

\bibitem[Lin \& Pringle(1987)]{lin87}
 Lin D.N.C., Pringle J.E., 1987, MNRAS, 225, 607

\bibitem[Lodato \& Rice(2004)]{lodato04}
 Lodato G., Rice W.K.M., 2004, 351, 630

\bibitem[Lodato \& Clarke(2011)]{lodato11}
 Lodato G., Clarke C.J., 2011, MNRAS, 413, 2735

\bibitem[Lodato \& Price(2010)]{lodato10}
 Lodato G., Price D., 2010, MNRAS, 405, 1212

\bibitem[Mayer et al.(2007)]{mayer07}
 Mayer L., Lufkin G., Quinn T., Wadsley J., 2007, ApJ, 661, L77

\bibitem[Meru \& Bate(2011)]{meru11}
 Meru F., Bate M.R., 2011, MNRAS, 411, L1

\bibitem[Monaghan(1992)]{monaghan92}
 Monaghan J.J., 1992, ARA\&A, 30, 543

\bibitem[Murray(1996)]{murray96}
 Murray J.R., 1996, MNRAS, 279, 402

\bibitem[Paardekooper, Baruteau \& Meru(2011)]{paardekooper11}
 Paardekooper S.-J., Baruteau C., Meru F., 2011, MNRAS, in press

\bibitem[Paczy\'nski(1978)]{paczynski78}
 Paczy\'nski B., 1978, Acta. Astron., 28, 91

\bibitem[Rafikov(2005)]{rafikov05}
 Rafikov R.R., 2005, AJ, 621, 69

\bibitem[Rice et al.(2011)]{rice11}
 Rice W.K.M., Armitage P.J., Mamatsashvili G., Lodato G., Clarke C.J., 2011, MNRAS, in press

\bibitem[Rice, Mayo \& Armitage (2010)]{rice10}
 Rice W.K.M., Mayo J.H., Armitage P.J., 2010, MNRAS, 402, 1740

\bibitem[Rice et al.(2003)]{rice03}
 Rice W.K.M., Armitage P.J., Bate M.R., Bonnell I.A., 2003, MNRAS, 339, 1025

\bibitem[Rice, Lodato \& Armitage(2005)]{rice05}
 Rice W.K.M., Lodato G., Armitage P.J., 2005, MNRAS, 364, L56

\bibitem[Ritchie \& Thomas(2001)]{ritchie01}
 Ritchie B.W., Thomas P.A., 2001, MNRAS, 323, 743

\bibitem[Shakura \& Sunyaev(1973)]{shakura73}
 Shakura N.I., Sunyaev R.A., 1973, A\&A, 24, 337

\bibitem[Shlosman \& Begelman(1989)]{shlosman89}
 Shlosman I., Begelman M., 1989, ApJ, 341, 685

\bibitem[Stamatellos \& Whitworth(2008)]{stamatellos08}
 Stamatellos D., Whitworth A.P., 2008, A\&A, 480, 879

\bibitem[Stamatellos et al.(2007)]{stamatellos07}
 Stamatellos D., Whitworth A.P., Bisbas T., Goodwin S., 2007, A\&A, 475, 37

\bibitem[Toomre(1964)]{toomre64}
 Toomre A., 1964, ApJ, 139, 1217

\bibitem[Truelove et al.(1997)]{truelove97}
 Truelove J.K., Klein R.I., McKee C.F., Holliman J.H., Howell L.H., Greenhough J.A., 1997, ApJ, 489, L179

\bibitem[Whitehouse \& Bate(2004)]{whitehouse04}
 Whitehouse S.C., Bate M.R., 2004, MNRAS, 353, 1078

\end{thebibliography}
\end{document}